\documentclass[conference,a4paper]{IEEEtran}

\usepackage[utf8]{inputenc} 
\usepackage[T1]{fontenc}
\usepackage{array}
\usepackage{url}
\usepackage{ifthen}
\usepackage{cite}
\usepackage{amssymb}
\usepackage{amsfonts}
\usepackage{dsfont}
\usepackage{enumitem}
\usepackage{tabularx}
\usepackage{graphicx}
\usepackage{xfrac}
\usepackage[cmex10]{amsmath} 

\newtheorem{theorem}{Theorem}

\newtheorem{corollary}[theorem]{Corollary}

\newcommand{\argmin}{\mathrm{arg}\displaystyle\min}

\newcommand{\qed}{\nobreak \ifvmode \relax \else
      \ifdim\lastskip<1.5em \hskip-\lastskip
      \hskip1.5em plus0em minus0.5em \fi \nobreak
      \vrule height0.75em width0.5em depth0.25em\fi}

\newcommand{\RNum}[1]{\uppercase\expandafter{\romannumeral #1\relax}}

\newcolumntype{M}[1]{>{\centering\arraybackslash}m{#1}}
\newcolumntype{N}{@{}m{0pt}@{}}

\DeclareFontFamily{OT1}{pzc}{}
\DeclareFontShape{OT1}{pzc}{m}{it}{<-> s * [1.10] pzcmi7t}{}
\DeclareMathAlphabet{\mathpzc}{OT1}{pzc}{m}{it}

\begin{document}
\title{On the Universality of the Logistic Loss Function} 


\author{%
  \IEEEauthorblockN{Amichai Painsky and Gregory Wornell}
  \IEEEauthorblockA{EECS Department, MIT,\\
                    Cambridge, MA 02139,\\
                    Email: \{amichai, gww\}@mit.edu}
}


\maketitle

\begin{abstract}
  A loss function measures the discrepancy between the true values (observations) and their estimated fits, for a given instance of data. A loss function is said to be proper (unbiased, Fisher consistent) if the fits are defined over a unit simplex, and the minimizer of the expected loss is the true underlying probability of the data. Typical examples are the zero-one loss, the quadratic loss and the Bernoulli log-likelihood loss (log-loss). In this work we show that for binary classification problems, the divergence associated with smooth, proper and convex loss functions is bounded from above by the Kullback-Leibler (KL) divergence, up to a multiplicative normalization constant. It implies that by minimizing the log-loss (associated with the KL divergence), we minimize an upper bound to any choice of loss functions from this set. This property justifies the broad use of log-loss in regression, decision trees, deep neural networks and many other applications. In addition, we show that the KL divergence bounds from above any separable Bregman divergence that is convex in its second argument (up to a multiplicative normalization constant). This result introduces a new set of divergence inequalities, similar to the well-known Pinsker inequality. 
\end{abstract}


\section{Introduction}
\label{intro}

Consider a weather forecaster that estimates the probability of rain on the following day. Its performance may be evaluated by different statistical measures. For example, we may count the number of times it assessed the chance of rain as greater than $t=50\%$, while it eventually did not rain (and vice versa). This corresponds to a $0$-$1$ loss (Table \ref{table:loss_functions}). Alternatively, we may choose different threshold values, $t$, or completely different measures (quadratic loss, Bernoulli log-likelihood loss, etc.).  Choosing a ``good" measure is a well-studied problem, mostly in the context of \textit{scoring rules} in decision theory 
\cite{gneiting2007strictly}. Assuming that the desired measure is known in advance, the weather forecaster may be designed accordingly, to minimize that measure. In practice, different tasks entail inferring different information from the provided estimates. It means that the forecaster shall be designed according to a single measure that is ``suitable" for a variety of possible purposes. This requirement is obviously quite challenging.

In this work we address this problem, as we show that for binary classification, the Bernoulli log-likelihood loss (log-loss) is a ``universal" choice which dominates any alternative  ``analytically convenient" loss function (smooth, proper and convex). Specifically, we show that by minimizing the log-loss we minimize the \text{regret} (defined in Section \ref{basic}) associated with all possible alternatives in this set. This result justifies the use of log-loss in many learning applications, as it is the only measure that provides such universality guarantees. 

Over the years, the log-loss was shown to have several favorable properties (for example, \cite{cover2012elements,merhav1998universal, jiao2014information}). However, these properties are mostly motivated by information-theoretic principles. Here, we justify the use of log-loss directly from a decision theory perspective.

In addition, we show that our universality result may be viewed from a divergence analysis viewpoint, as we show that the divergence associated with the log-loss (KL divergence) bounds from above any separable Bregman divergence that is convex in its second argument, up to a multiplicative normalization constant. This result provides a new set of divergence inequalities, which have a similar nature as the well-known Pinsker inequality \cite{cover2012elements}. In that sense, our Bregman analysis may be viewed as a complementary set of results to the well known Pinsker-like $f$-divergence inequalities \cite{sason2016f}.

\section{Basic Definitions}
\label{basic}
Let $Y\in \{0,1\}$ be a Bernoulli distributed binary random variable with a parameter $p$. Let $\hat{Y}$ be an estimate of $Y$. A loss function $l(y,\hat{y})$ quantifies the difference between a realization of $Y$ and its corresponding estimate. In this work we focus on probabilistic estimates, for which $\hat{y} \triangleq q \in [0,1]$. In other words, $\hat{y} \triangleq q$ is a ``soft" decision that corresponds to the probability of the event $y=1$ (as opposed to a ''hard decision" in which $\hat{y} \in \{0,1\}$).  A Binary loss function is defined as
\begin{align}
l(y,q)=\mathds{1}\{y=0\}l_0(q)+\mathds{1}\{y=1\}l_1(q)    
\end{align}
where $\mathds{1}\{\cdot\}$ is an indicator function and $l_k(q)$ is a loss function associated with the event $y=k$. Several examples of  typical loss functions, such as $0$-$1$ loss, quadratic loss and others are provided in Table \ref{table:loss_functions}. Let 
\begin{align}
    L(p,q)=E_Yl(Y,q)
\end{align}
be the expected loss with respect to $Y$. Notice that $L(p,q)$ only depends on the Bernoulli parameter $p$ and the estimate $q$. A \textit{proper loss function} is a loss function for which the minimizer of the expected loss is the true underlying distribution of the random variable we are to estimate, 
$p=\argmin_q{L(p,q)}$. This property is also known as Fisher-consistent or unbiased loss. A \textit{strictly} proper loss function means that $q=p$ is a unique minimizer. In this work we require several regularity conditions for proper loss functions. We say that a proper loss function is \textit{fair} if $l_0(0)=l_1(1)=0$. This means that there is no loss incurred for perfect prediction. Further, We say that a proper loss function is \textit{regular} if $\lim_{q\searrow0}ql_1(q)=\lim_{q\nearrow1}(1-q)l_0(q)=0$. Intuitively, this condition ensures that making mistakes on events that never happen should not incur a penalty. In this paper we consider loss functions that are \textit{fair} and \textit{regular} unless stated otherwise.

Define the minimum of the expected proper loss as $G(p) \triangleq L(p,p)$.  This term is also  known as the \textit{generalized entropy function} \cite{gneiting2007strictly}, \textit{Bayes risk} \cite{reid2010composite} or \textit{Bayesian envelope} \cite{merhav1993universal}. For example, assuming $l(y,q)$ is the log-loss, then the generalized entropy function is Shannon entropy. Additional examples appear in Table \ref{table:loss_functions}. The \textit{regret} is defined as the difference between the expected loss and its minimum. For proper loss functions we have that $\Delta L(p,q)=L(p,q)-G(p)$. Savage \cite{savage1971elicitation} showed that a loss function  $l(y,q)$ is proper and regular iff $G(p)$ is concave and for every $p,q \in[0,1]$ we have that $$L(p,q)=G(q)+(p-q)G'(q).$$
This property allows us to draw an immediate connection between regret and Bregman divergence. Specifically, let  $f:\mathbb{S}\rightarrow \mathbb{R}$ be a convex function over some convex set $\mathbb{S}\in\mathbb{R}^n$. Then its associated Bregman divergence is defined as $$D_f(s||s_0)=f(s)-f(s_0)-\langle s-s_0,\nabla f(s_0) \rangle$$
for any $s,s_0 \in \mathbb{S}$, where $\nabla f(s_0)$ is the gradient of $f$ at $s_0$. By setting $s=[0,1]$ we have that $\nabla f = f'$ and $\Delta L(p,q) = D_{-G}(p,q)$. This means that the regret of a proper loss function is uniquely associated with a Bregman divergence. An important example is the Kullback-Leibler (KL) divergence, $D_{\mathrm{KL}}(p||q)$ associated with the log-loss. Additional examples appear in Table \ref{table:loss_functions}.  

Convex loss functions hold a special role in learning theory and optimization \cite{buja2005loss,reid2010composite}. Let $\underline{X}$ and $Y$ be a set of explanatory variables (features) and an independent variable (target) respectively. Given a set of $n$ i.i.d. samples of $\underline{X}$ and $Y$, the empirical risk minimization (ERM) criterion seeks to minimize
$\frac{1}{n}\sum_{i=1}^n \mathds{1}(y_i=0)l_0(q_i)+\mathds{1}(y_i=1)l_1(q_i)$, 
where $q_i \triangleq q_i(\underline{x}_i)$. As the complexity of this problem increases, it is desirable for this minimization problem to be convex in the optimization parameter. Alternatively, assuming that $p$ is known,  minimizing the expected loss $L(p,q)$ has many desirable properties, both analytically and computationally, when the problem is convex. It is important to mention that convex proper loss functions correspond to Bregman divergences that are convex in their second parameter. This family of divergences are of a special interest in many applications \cite{bauschke2001joint,byrne2014iterative}, and have an important role in our results.

\section{Main Result}
Our main result is as follows,

\begin{theorem}
\label{main_theorem}
Let $l(y,q)$ be a smooth and proper binary loss function with a corresponding generalized entropy function $G$. Assume that $l(y,q)$  is convex in $q$. Then for every $p,q \in [0,1]$,
$$D_{\mathrm{KL}}(p||q)\geq \frac {1}{C(G)} D_{-G}(p||q)   $$
where  $C(G)>-\frac{1}{2}G''(p)|_{p=\frac{1}{2}}$ is a normalization constant (that does not depend on $p$ or $q$).
\end{theorem}
A proof of this theorem is provided in Appendix A. 

This result established that the KL divergence, associated with the log-loss, bounds from above the divergence of any smooth, proper and convex loss function, up to a multiplicative constant. In other words, by minimizing the log-loss we minimize an upper bound on any choice of such loss functions. The practical implications of this result are quite immediate. Assume that the performance measure according to which a learning algorithm is to be measured with is unknown a-priori to the experiment (for example, the weather forecaster, discussed in Section \ref{intro}). Then, minimizing the log-loss provides an upper bound to any possible choice of measure, associated with an ``analytically convenient" loss function. This property makes the log-loss a universal choice for classification problems as it governs a large and significant class of measures.  

We notice that the normalization constant $C(G)$ seems to provide a bound that is untight. However, it is unavoidable since proper loss functions are closed under affine transformations. 
It is important to mention that typical minimal values of $C(G)$ are relatively ``small". For example, we have that $C(G)=1$ for $0$-$1$ loss and $C(G)=2$ for both the quadratic loss and Boosting loss \cite{buja2005loss}. The corresponding quadratic bound, for example,  is $D_{\mathrm{KL}}(p||q)\geq(p-q)^2$. 

In addition to the universality property, Theorem \ref{main_theorem} allows us to analyze the local behavior of divergences associated with smooth, proper and convex loss functions, as demonstrated in Corollary \ref{local_analysis}.

\begin{corollary}
Let $l(y,q)$ be a smooth and proper binary loss function with a corresponding generalized entropy function $G$. Assume that  $l(y,q)$ is convex in $q$. Then for every $p,p+dp \in [0,1]$, 
$$\frac{1}{C(G)}D_{-G}(p||p+dp) \lesssim \frac{dp^2}{2} \mathcal{I}(p)   $$
where $\mathcal{I}(p)$ is the Fisher information of a Bernoulli distributed random variable with a parameter $p$, and  $\lesssim$  refers to inequality up to second order of Taylor expansion terms, $O(dp^2)$. 
\label{local_analysis}
\end{corollary}
A proof of this theorem is provided in Appendix B.

Corollary \ref{local_analysis} implies that when $q$ is ``close enough" to $p$, the divergence associated with the set of smooth, proper and convex binary loss functions is governed by the Fisher information of a Bernoulli random variable with a parameter $p$ (up to the second order terms of the Taylor expansion). Since $\mathcal{I}(p)$ corresponds to the KL divergence (and henceforth, to log-loss) we conclude that the rate of convergence of any  $D_{-G}(p||q)$ in this set is bounded from above by the rate of $D_{\mathrm{KL}}(p||q)$, when $q=p+dp$. This provides an interesting trade-off between the universality of the log-loss, and its slower rate of convergence. 

As stated in Section \ref{basic}, the divergence associated with a convex and proper loss function is a Bregman divergence that is convex in its second argument. This allows us to present our results from a divergence analysis perspective. Further, it allows us the extend our study to a greater alphabet size. Define a \textit{separable} Bregman divergence as 
$$D_g(\underline{p}||\underline{q})=\sum_{i=1}^{m}g(p_i)-g(q_i)-g'(q_i)(p_i-q_i) $$
for any $\underline{p},\underline{q}\in [0,1]^m$ and a convex function $g:[0,1]\rightarrow[0,\infty]$. Notice that in the general case, the Bregman divergence is not restricted to the unit simplex. Separable Bregman divergences hold a fundamental role in divergence analysis, as shown in \cite{harremoes2007information,jiao2014information}.
\begin{theorem}
\label{divergence_theorem}
Let $D_g(\underline{p}||\underline{q})$ be a separable Bregman divergence, that is convex in $\underline{q}$. Then, for every $\underline{p}, \underline{q} \in [0,1]^m$, 
$$D_{\mathrm{KL}}(\underline{p}||\underline{q}) \geq \frac{1}{C(g)}D_g(\underline{p}||\underline{q})$$
where $D_{\mathrm{KL}}(\underline{p}||\underline{q})=\sum_{i=1}^m p_i \log \frac{p_i}{q_i} -\sum_{i=1}^m p_i +\sum_{i=1}^m q_i$ and $C(g)>g''(p)|_{p=1}$ is a normalization constant.
\end{theorem}
As a corollary, we further show that Theorem \ref{divergence_theorem} holds for the case where $\underline{p}$ and $\underline{q}$ are constrained to the unit simplex. Proofs of Theorem \ref{divergence_theorem} and its corollary are provided in Appendix C. 

As in Theorem \ref{main_theorem}, we notice that the constant $C(g)$ is unavoidable since affine transformations preserve the convexity of $D_g(\underline{p},\underline{q})$. For example, for the squared error case we have that 
\begin{align}
\label{mse}
    D_{\mathrm{KL}}(\underline{p}||\underline{q}) \geq \frac{1}{2} \sum_{i=1}^m (p_i-q_i)^2.
\end{align} Notice that the multiplicative constant in (\ref{mse}) is different than in the binary case ($C(g)=1$). This is a result of the fundamental difference between the optimally conditions when minimizing some function $f(p_1,p_2)$ under the constraint that $p_1+p_2=1$, as opposed to minimizing the same function using a single parameter,  $f(p_1,1-p_1)$.

Further, it is important to mention that (\ref{mse}) resembles the well-known Pinsker inequality \cite{cover2012elements}, that states 
\begin{align}
\label{pinsker}
    D_{\mathrm{KL}}(\underline{p}||\underline{q}) \geq \frac{1}{2} \left(\sum_{i=1}^m |p_i-q_i|\right)^2.
\end{align}
Notice that the right-hand side of (\ref{pinsker}) is not a Bregman divergence (in fact it is a squared Csisz{\'a}r divergence \cite{cover2012elements}) and therefore it is not considered in Theorem \ref{divergence_theorem}.

It is easy to verify that the Pinsker inequality is tighter than (\ref{mse}). However, the squared-error bound (\ref{mse}) is just a simple case of our broader result. In this sense,  Theorem \ref{divergence_theorem} may be viewed as an extension of Pinsker-like inequalities to the family of separable Bregman divergences that are convex in their second argument.

\begin{table*}[!ht]
\caption{Examples of binary loss functions}
\renewcommand{\baselinestretch}{1}\footnotesize
\label{table:loss_functions} 
\centering
\begin{tabular}{|M{2cm}|M{2.8cm}|M{2.5cm}|M{4cm}|M{2.5cm}|N}  

\hline

Loss function
& $l(y,q)$
& $G(p)=L(p,p)$ 
& $D_{-G}(p||q)$ 
& $w(p)$ &\\[15pt]

\hline
\hline

\begin{tabular}{@{}c@{}} 0-1 loss\end{tabular}  
& \begin{tabular}{@{}c@{}} $y\mathds{1}\{q<\frac{1}{2}\}+ $ \quad\quad \\ $(1-y)\mathds{1}\{q \geq \frac{1}{2}\}$\end{tabular} 
& \begin{tabular}{@{}c@{}} $p\mathds{1}\{p<\frac{1}{2}\}+ \quad$ \\ $(1-p)\mathds{1}\{p \geq \frac{1}{2}\}$\end{tabular} 
&
\begin{tabular}{@{}c@{}} $(1-2p)\mathds{1}\{p<\frac{1}{2},q\geq\frac{1}{2}\}+$ \\ $(2p-1)\mathds{1}\{p \geq \frac{1}{2},q<\frac{1}{2}\} $  \;\; \end{tabular} 
&
\begin{tabular}{@{}c@{}} $2\delta(\frac{1}{2}-p)$ \end{tabular}
&\\[22pt] \hline

\begin{tabular}{@{}c@{}} Quadratic loss\end{tabular}  
& \begin{tabular}{@{}c@{}} $y(1-q)^2+(1-y)q^2$\end{tabular} 
& \begin{tabular}{@{}c@{}} $p(1-p)$\end{tabular} 
&
\begin{tabular}{@{}c@{}} $(p-q)^2 $  \;\; \end{tabular} 
&
\begin{tabular}{@{}c@{}} $2$ \end{tabular}
&\\[22pt] \hline

\begin{tabular}{@{}c@{}} Log loss\end{tabular}  
& \begin{tabular}{@{}c@{}} $y\log \frac{1}{q}+ $ \quad\quad\quad \\ $(1-y)\log \frac{1}{1-q}$\end{tabular} 
& \begin{tabular}{@{}c@{}} $p\log \frac{1}{p}+ $ \quad\quad\quad \\ $(1-p)\log \frac{1}{1-p}$\end{tabular} 
&
\begin{tabular}{@{}c@{}} $p \log \frac{p}{q} + (1-p)\log \frac{1-p}{1-q} $  \;\; \end{tabular} 
&
\begin{tabular}{@{}c@{}} $\frac{1}{p(1-p)}$ \end{tabular}
&\\[24pt] \hline

\begin{tabular}{@{}c@{}} Boosting loss\end{tabular}  
& \begin{tabular}{@{}c@{}} $2y\sqrt{\frac{1-q}{q}}+ $ \quad\quad\quad \\ $2(1-y)\sqrt{\frac{q}{1-q}}$\end{tabular} 
& \begin{tabular}{@{}c@{}} $4\sqrt{p(1-p)}$\end{tabular} 
& \begin{tabular}{@{}c@{}} $2\left( p\sqrt{\frac{1-q}{q}}+(1-p)\sqrt{\frac{q}{1-q}}\right)-$ \quad\quad\quad \\ $4\sqrt{p(1-p)}$\end{tabular} 
&
\begin{tabular}{@{}c@{}} $\frac{1}{(p(1-p))^{\sfrac{3}{2}}}$ \end{tabular}
&\\[24pt] \hline

\end{tabular}
\end{table*}

\section{Illustrations}
\begin{figure}[ht]
\centering
\includegraphics[width =0.4\textwidth,trim={2cm 7cm 2cm 7.9cm}]{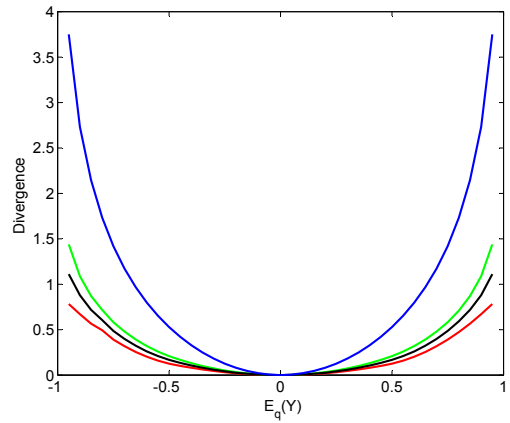}
\caption{Fixed expectation experiment: $Y\in\{-1,0,1\}$, $p=\left[ \sfrac{1}{4},\sfrac{1}{2},\sfrac{1}{4}\right]^T$ and we optimize over $q$ under a fixed expectation constraint, $E_q(Y)$. The blue (upper) curve  is $D_{\mathrm{KL}}(\underline{p}||\underline{q})$ while the rest of the curves are Bregman divergences, that are convex in $\underline{q}$.}
\label{fixed_stat_experiment}
\end{figure}

We demonstrate our results in two illustrative experiments. In the first experiment we focus on ternary alphabet $y\in\{-1,0,1\}$ with a corresponding distribution $p(y)=\left[ \sfrac{1}{4},\sfrac{1}{2},\sfrac{1}{4}\right]^T$. We are interested in $q$ such that $E_q(Y)$ is a given fixed value. This implies a fixed expectation constraint on $q$. We examine smooth, proper and convex loss functions through their corresponding Bregman divergences (that are convex in their second moment), as discussed above. Notice that our constraint is linear in the $q$, so that our optimization problem is convex minimization over a convex set. Notice this problem may be easily generalized to larger alphabets and additional constraints on greater moments of $Y$. Fig. \ref{fixed_stat_experiment} demonstrates the results we achieve. The blue (upper) curve is $D_{\mathrm{KL}}(\underline{p}||\underline{q})$ and the rest of the curves correspond to different Bregman divergences 
including quadratic loss and separable Mahalanobis distances \cite{gneiting2007strictly}.  
We first notice that $E_p(Y)=0$. This means that by allowing $E_q(Y)=0$, we get an unbiased estimate and the divergence is zero. On the other hand, different values of $E_q(Y)$ result in bias. In this case, it is evident that the KL divergence bounds from above any choice of divergence, as expected. 

\begin{figure}[ht]
\centering
\includegraphics[width =0.47\textwidth,trim={2cm 7cm 2cm 7.2cm}]{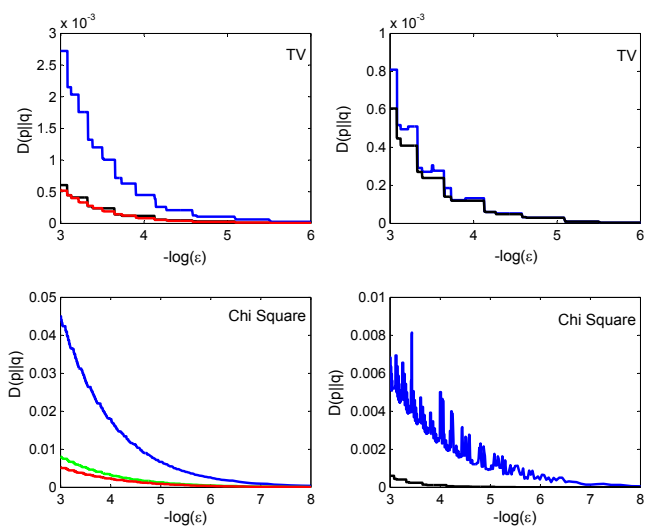}
\caption{Divergence constraint experiment: we minimize $D_g(\underline{p}||\underline{q})$ under the constraint that $D_{\epsilon}(\underline{p}||\underline{q})\geq \epsilon$ for $D_{\epsilon}(\underline{p}||\underline{q})$ being total variation (upper charts) and Chi Square (lower charts). $D_g(\underline{p}||\underline{q})$ are the same as in Fig. \ref{fixed_stat_experiment}. The charts on the left demonstrate Theorem \ref{divergence_theorem} and the charts on the right demonstrate inequality (\ref{experiments_inequality}), as described in the main text.}
\label{epsilon_experiment}
\end{figure}

In the second experiment we show that our bound holds for a broader range of practical problems. Assume that there exists some constraint that prevents $\underline{q}$ from converging to $\underline{p}$. This constraint may be statistical, computational or even algorithmic. We model this problem by stating that $D_{\epsilon}(\underline{p}||\underline{q})\geq \epsilon$ for some (unknown) divergence measure $D_\epsilon$ and $\epsilon>0$. In other words, we restrict $\underline{q}$ to be $\epsilon$-far from $\underline{p}$ (in a $D_{\epsilon}(\underline{p}||\underline{q})$ sense). Fig. \ref{epsilon_experiment} demonstrates the results we achieve for $D_{\epsilon}(\underline{p}||\underline{q})=\sum_{i=1}^m |p_i-q_i|$ (total variation, in the upper charts) and $D_{\epsilon}(\underline{p}||\underline{q})=\sum_{i=1}^m \frac{(p_i-q_i)^2}{q_i}$ (Chi-Square, in the lower charts). The charts on the left show different divergence measures (in the same manner as in the previous experiment) for different $\epsilon$ values. The charts on the right demonstrate the KL divergence (blue curve on top), and the quadratic divergence, when we plug the minimizer of the KL divergence. Specifically, $\sum_{i=1}^m p_i\log \frac{p_i}{q^{\mathrm{KL}}_i}$ and $\sum_{i=1}^m(p_i-q^{\mathrm{KL}}_i)^2$ where $\underline{q}^{\mathrm{KL}}=\argmin_{\underline{q}}  D_{\mathrm{KL}}(\underline{p}||\underline{q})$.

We first notice that our bound holds for the two choices of $D_{\epsilon}(\underline{p}||\underline{q})$, where greater $\epsilon$ values result in a greater bias than lower values, as expected. Second, notice that 
\begin{align}
\label{experiments_inequality}
D_{\mathrm{KL}}(\underline{p}||\underline{q})\geq D_{\mathrm{KL}}(\underline{p}||\underline{q}^{\mathrm{KL}}) \geq \frac{1}{C(g)} D_{g}(\underline{p}||\underline{q}^{\mathrm{KL}}) 
\end{align}
for any separable Bregman divergence that is convex in $\underline{q}$. This means that we may use the minimizer of the KL divergence as an (untight) approximated ``solution" for $D_{g}(\underline{p}||\underline{q})$. Indeed, the charts on the left demonstrate this inequality for quadratic divergence.   

\section{Conclusions and discussion}
In this work we introduce a fundamental inequality for divergence measures associated with smooth, proper and convex binary loss functions. We show that the KL divergence, associated with the Bernoulli log-likelihood loss function, bounds from above any divergence associated with this set of losses. This property makes the log-loss a universal choice, in the sense that it controls any ``analytically convenient" alternative one may be interested in. The implications of this result span a broad variety of applications. 
In binary classification trees, the split criterion in each node is typically chosen between the Gini impurity (which corresponds to quadratic loss) and Information-Gain (corresponds to log-loss). The choice of a suitable splitting mechanism holds a long standing discussion with many statistical and computational implications (for example, \cite{painsky2017cross}). In deep neural networks, the objective function is cross-entropy minimization (which again corresponds to log-loss) where several alternatives have (empirically) shown to be less successful over the years. Further, our result may extend the fundamental PAC-Bayes bound \cite{langford2005tutorial} to a universal setup which is independent in the choice of the loss.

As demonstrated in Theorem \ref{divergence_theorem}, our results may be viewed from a Bregman divergences perspective. Here, the applications of our results are universality guarantees for distributional clustering \cite{pereira1993distributional}, clustering with Bregman divergences \cite{banerjee2005clustering} and many others.

\section*{Appendix A: Sketch of proof for Theorem \ref{main_theorem}}
A smooth and proper binary loss function satisfies
$$\frac{\partial}{\partial q} L(p,q)|_{q=p}=pl'_0(p)+(1-p)l'_1(p)=0.$$
This means that 
$$\frac{-l'_1(p)}{1-p}=\frac{l'_0(p)}{p}  \triangleq w(p)   $$
where $w(p)$ is defined as the \textit{weight function}. Shuford et at. \cite{shuford1966admissible} showed that the converse is also true: a smooth binary loss function is proper only if the above holds, for $w(p)$ that satisfies $\int_\epsilon^{1-\epsilon}w(c)dc < \infty$, for all $\epsilon>0$. Typical examples of weight functions for different losses appear in Table \ref{table:loss_functions}. In addition, it is easy to verify that $\frac{d^2}{dp^2}G(p)=-w(p)$ for all proper binary loss functions. The convexity of the loss (with respect to $q$) implies that 
$$\frac{\partial^2}{\partial q^2} L(p,q) = w(q)+(q-p)w'(q)\geq 0 $$ for every fixed $p\in[0,1]$. Plugging $p=0$ and $p=1$ we get
$$-\frac{1}{q}\leq \frac{w'(q)}{w(q)}\leq\frac{1}{1-q}  $$ for all $q\in (0,1)$. Integrating both sides achieves 
\begin{align}
\label{basic_inq}
&\text{For}\;\;  1>q\geq \frac{1}{2}: \quad \frac{w\left(\frac{1}{2}\right)}{2q}\leq w(q) \leq    \frac{w\left(\frac{1}{2}\right)}{2(1-q)}\quad\quad\quad\quad\quad\\\nonumber
&\text{For}\;\; \frac{1}{2}>q>0 : \quad \frac{w\left(\frac{1}{2}\right)}{2q}\geq w(q) \geq    \frac{w\left(\frac{1}{2}\right)}{2(1-q)}.\quad\quad
\end{align}
Similar results appear in Theorem 29 of \cite{reid2010composite}. 
Let us now look at $R(p,q)=C \cdot D_{\mathrm{KL}}(p||q)-D_{-G}(p||q)$ for a fixed $p$ and find such $C$ for which $R\geq0$ for all $p,q$. We have 
\newline
\begin{description}[labelindent=0cm]\itemsep6pt
\item $\frac{\partial}{\partial q} R(p,q)=(q-p)\left(\frac{C}{q(1-q)}-w(q)\right)$
\item $\frac{\partial^2}{\partial q^2} R(p,q)=C\left(\frac{p}{q^2}+\frac{1-p}{(1-q)^2}\right)-w(q)-(q-p)w'(q).$
\end{description}
\hfill \break
We require that $q=p$ is a minimum. Notice that the second derivative condition, together with (\ref{basic_inq}), yield that
$C>\frac{1}{2}w\left(\frac{1}{2}\right)=-\frac{1}{2}G''(p)|_{q=p}$. This further implies that $\frac{C}{q(1-q)}-w(q)>0$ so that $q=p$ is the global minimum (according to the first derivative condition), as desired. \hfill $\square$

\section*{Appendix B: Sketch of proof for Corollary \ref{local_analysis}}
Assume that $p,p+dp \in [0,1]$. Then, we may derive the Taylor expansion of $D_{-G}(p||p+dp)$ around $p$:
\begin{align}
D_{-G}(p||p+dp)=&\frac{dp^2}{2}\frac{d^2}{dp^2}D_{-G}(p||q)|_{q=p}+O(dp^2) \simeq \\\nonumber &\frac{dp^2}{2}\frac{d^2}{dp^2}L(p,q)|_{q=p}=\frac{dp^2}{2}w(p)<\\\nonumber
&\frac{dp^2}{2}\frac{C}{p(1-p)}=\frac{dp^2}{2}\cdot C \cdot \mathcal{I}(p). \quad \quad \quad    \square
\end{align} 

\section*{Appendix C: Sketch of proof for Theorem \ref{divergence_theorem}}
Let us begin by stating the derivatives of a separable Bregman divergence,
\newline
\begin{description}[labelindent=0cm]\itemsep6pt
\item $\frac{\partial}{\partial q_i} D_g(\underline{p}||\underline{q})=g''(q_i)(p_i-q_i)$
\item $\frac{\partial^2}{\partial q_i^2} D_g(\underline{p}||\underline{q})=g'''(q_i)(q_i-p_i)+g''(q_i).$
\end{description}
\hfill \break
Assuming a fixed \underline{p}, the convexity of $D_g(\underline{p}||\underline{q})$ implies that its second derivative is non-negative for every $p_i\in[0,1]$. Specifically, for $p_i=\{0,1\}$  we get that 
$$-\frac{1}{q_i} \leq \frac{g'''(q_i)}{g''(q_i)}\leq \frac{1}{1-q_i}.$$ 
Integrating the left inequality with respect to $q_i$ attains
\begin{align}
\label{appendix_C_inequality}
    g''(q_i) \leq \frac{g''(1)}{q_i}.
\end{align}
As in Appendix A, we define $R(\underline{p},\underline{q})=C\cdot D_{\mathrm{KL}}(\underline{p}||\underline{q})-D_g(\underline{p}||\underline{q})$ and show that it is non-negative. Let us fix $\underline{q}$ and analyze $R(\underline{p},\underline{q})$ with respect to $\underline{p}$. We have that 
\newline
\begin{description}[labelindent=0cm]\itemsep6pt
\item $\frac{\partial}{\partial p_i} R(\underline{p},\underline{q})=C\left(\log \frac{p_i}{q_i} \right)-g'(p_i)+g'(q_i)$
\item $\frac{\partial^2}{\partial p_i^2} R(\underline{p},\underline{q})=\frac{C}{p_i}-g''(p_i).$
\end{description}
\hfill \break
Notice that the second derivative, together with (\ref{appendix_C_inequality}) implies that $R(\underline{p},\underline{q})$ is convex (in $\underline{p}$) if $C\geq g''(1)$. This further means that $p_i=q_i$ is a unique minimizer, for this choice of $C$. Repeating the same derivation for any fixed $\underline{q}$ yields the desired property. 

Assume now that $\underline{p}$ is given and it is over the unit simplex, while $\underline{q}$ is constrained to the same domain. Then, we may repeat the derivation above with corresponding Lagrange multipliers and reattain condition (\ref{appendix_C_inequality}). Further, we may show that in this case, $R(\underline{p},\underline{q})\geq0$ for every $\underline{p},\underline{q}\in[0,1]^m$. This means that our inequality hold for any subset of $[0,1]^m$, such as the unit simplex, for example. \hfill $\square$



\bibliographystyle{IEEEtran}
\bibliography{bibi}

\end{document}